\documentclass{WileyMSP-template}
\usepackage{amsmath,graphicx,array}
\usepackage{dcolumn,soul}%
\usepackage{amsthm}
\usepackage[figuresright]{rotating}%
\usepackage{algorithm, algorithmicx, algpseudocode}
\usepackage{listings}%
\usepackage{hyperref}
\usepackage{textcomp}
\usepackage{gensymb}
\usepackage{xcolor}
\usepackage{graphicx}
\usepackage[square,comma,super,sort&compress]{natbib}
\begin{document}

\pagestyle{fancy}
\rhead{\includegraphics[width=2.5cm]{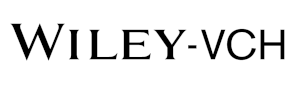}}

\title{Quantitatively predicting angle-resolved polarized Raman intensity of anisotropic layered materials}
\maketitle


\author{Jia-Liang Xie, Tao Liu, Yu-Chen Leng, Rui Mei, Heng Wu, Chen-Kai Liu, Jia-Hong Wang, Yang Li, Xue-Feng Yu,}
\author{Miao-Ling Lin,$^{*}$ and}
\author{Ping-Heng Tan$^{*}$}

\dedication{}

\begin{affiliations}
J.-L. Xie, T. Liu, Y.-C. Leng, R. Mei, H. Wu, C.-K. Liu, M.-L. Lin, P.-H. Tan\\
State Key Laboratory of Semiconductor Physics and Chip Technologies\\ Institute of Semiconductors\\ Chinese Academy of Sciences\\ Beijing 100083, China\\
E-mail: linmiaoling@semi.ac.cn; phtan@semi.ac.cn

J.-L. Xie, T. Liu, R. Mei, H. Wu, C.-K. Liu, J.-H. Wang, X.-F. Yu, M.-L. Lin, P.-H. Tan\\
Center of Materials Science and Optoelectronics Engineering\\ University of Chinese Academy of Sciences\\ Beijing 100049, China

J.-H. Wang, Y. Li, X.-F. Yu\\
Shenzhen Engineering Center for the Fabrication of Two-Dimensional Atomic Crystals\\ Shenzhen Institutes of Advanced Technology\\ Chinese Academy of Sciences\\ Shenzhen 518055, China

\end{affiliations}



\justifying
\begin{abstract}
  Angle-resolved polarized Raman (ARPR) spectroscopy provides insights into optical anisotropy and symmetry-related electron-photon/electron-phonon couplings of anisotropic layered materials (ALMs). However, since their discovery over ten years ago, ARPR responses in ALM flakes has exhibited a puzzling dependence on flake thickness, excitation wavelength, and dielectric environment, complicating their understanding and prediction. By taking black phosphorus (BP) ($\geq$20 nm) flakes and four-layer Td-WTe$_2$ as examples, this study introduces intrinsic Raman tensors ($\textit{\textbf{R}}^{\rm int}$) and proposes strategies to predict the ARPR intensity profiles of thick and atomically-thin ALM flakes by considering birefringence, linear dichroism and multilayer interference inside multilayered structures with experimentally determined complex refractive indexes along in-plane axes and complex tensor elements of $\textit{\textbf{R}}^{\rm int}$ for the corresponding phonon modes. The tensor elements of effective Raman tensors ($\textit{\textbf{R}}^{\rm eff}$), which are directly linked to the polarization vectors of incident and scattered light outside the ALM surface, were derived to quantitatively predict ARPR intensity for these ALM flakes, showing intricate dependence on ALM thickness, dielectric substrates, and excitation wavelengths. This framework can be extended to other ALM flakes from atomically-thin layers to bulk limit, facilitating comprehensive prediction of their ARPR intensity regardless of layer-dependent electronic properties.

\end{abstract}

\section{Introduction}
Anisotropic layered materials (ALMs) emerge with distinctive direction-selected characteristics due to their low-symmetry in crystal structures.\cite{Ma-2018-Nature, Biswas-2021-Science, Kim-2021-Nautre, Li-2022-AM, Luo-2023-NN, Tang-2023-NE, Chen-2023-NM, Huang-2024-Science} ALMs, e.g., black phoshporus (BP), ReX$_2$ (X = S, Se), GeSe$_2$ and Td-WTe$_2$, exhibit polarization-sensitive optical responses and significant anisotropy in phonon-related properties,\cite{Qiao-2014-NC,Li-2023-NC, Zhang-2025-NC, Yang-2018-JACS, Chen-2019-AM, Ho-2019-NanoEnergy, Mei-2023-AM} making them promising candidates for polarization-sensitive photodetectors,\cite{Zhang-2015-AFM,Yang-2019-AFM,Kim-2021-Small,Yuan-2021-NPht} field effect transistors\cite{Zhang-2015-AFM, Yang-2016-AM} and thermoelectric devices.\cite{Fei-2014-NanoLetter,Yang-2022-AEM,Pan-2022-NC,Bang-2023-IJER} Additionally, atomically-thin ALM flakes demonstrate pronounced layer-number-dependent electronic properties,\cite{Luo-2015-NC,Zhang-2018-SciAdv,Hwangbo-2021-NN,Qiao-2014-NC,Zhao-2021-NC,Song-2022-Nature,Zhang-2025-SciAdv} electron-phonon (e-phn)/electron-photon (e-pht)\cite{Luo-2015-NC,Zhang-2018-SciAdv,Hwangbo-2021-NN,Song-2022-Nature,Zhang-2025-SciAdv} as well as optical anisotropy,\cite{Qiao-2014-NC,Zhao-2021-NC,Song-2022-Nature} whereas thick ALM flakes showing bulk-like electronic band structure on dielectric substrates can generate optical cavity to enhance light-matter interactions and optical anisotropy.\cite{Zhang-np-2022} These characteristics further enable enhanced control over optical anisotropy and open up new possibilities for polarization-sensitive optoelectronic applications.\cite{Biswas-2021-Science,Liu-nc-2025} A comprehensive understanding of the intrinsic anisotropy effects in e-pht and e-phn interactions of ALM flakes from atomically-thin layers to the bulk limit is essential for advancing their applications in optoelectronics and leveraging their unique properties.\cite{Novko-2020-PRL,mao-jacs-2019,Lin-2020-SciBull}

Angle-resolved polarized Raman (ARPR) spectroscopy is a ubiquitous tool for comprehensively studying optical anisotropy, e-pht and e-phn couplings of ALMs,\cite{Chaudhary-2019-SciAdv,Lin-2020-SciBull,Choi-2020-NM,Kim-SciAdv,Pogna-2024-NC,Wang-2024-NN} which is obtained through altering the polarization vectors of incident laser ($\textit{\textbf{e}}_{\rm i}$) and scattered Raman signal ($\textit{\textbf{e}}_{\rm s}$) external to the material surface.\cite{Loudon-book-1964,strach-prb-1998,Lin-2020-SciBull} In contrast to in-plane isotropic materials where ARPR intensity ($I$) can be predicted by the crystal-symmetry defined Raman tensor (\textit{\textbf{R}}),\cite{Loudon-book-1964} i.e.,

\begin{equation}
 I\propto|\textit{\textbf{e}}^{\rm T}_{\rm s}\cdot \textit{\textbf{R}}\cdot \textit{\textbf{e}}_{\rm i}|^2,
\label{intensity1}
\end{equation}

\noindent the ARPR intensity of ALMs presents anomalous dependence on ALM flake thickness ($d_{ALM}$) and excitation wavelength ($\lambda_{\rm i}$), which has remained
unresolved since its discovery over a decade ago.\cite{Ribeiro-ACSNano-2015, Kim-nanoscale-2015, Mao-small-2016,Ling-Nanolett-2016,Heureux-nl-2016} This inspires huge efforts to understand the corresponding results by taking birefringence or linear dichroism effects into account.\cite{Ribeiro-ACSNano-2015, Kim-nanoscale-2015,Lin-2020-SciBull,Mao-small-2016,Ling-Nanolett-2016,Heureux-nl-2016,Zheng-pr-2018,Choi2020NH,Zou-2023-Small} In principle, birefringence and linear dichroism effects in ALMs should result in depth-dependent polarization and amplitude of both excitation and scattered electric fields at the location of the scattering event,\cite{Lin-2020-SciBull} which cannot be treated as constants of $\textit{\textbf{e}}_{\rm s}$ and $\textit{\textbf{e}}_{\rm i}$. Furthermore, for thick ALM flake, flake-substrate multilayer dielectrics can further modulate the light propagation within ALM flakes due to multilayer interference effects of excitation/scattered light.\cite{Zhang-np-2022} Nevertheless, due to the complicated depth-dependent polarization and intensity of excitation/scattered light inside flake-substrate multilayer dielectrics, the ARPR intensity is usually estimated by considering $\textit{\textbf{e}}_{\rm s}$ and $\textit{\textbf{e}}_{\rm i}$ external to the ALM surface and a Raman tensor with fitted complex elements.\cite{Ribeiro-ACSNano-2015, Kim-nanoscale-2015,Mao-small-2016,Ling-Nanolett-2016,Zheng-pr-2018,Choi2020NH,Zhu-ph-2020,ding-jpcc-2020,Zou-2023-Small} These fitted complex elements in \textit{\textbf{R}} should depend strongly on $d_{ALM}$ and $\lambda_{\rm i}$, challenging the established view that Raman tensor in thick flake with bulk-like electronic structure is an intrinsic material property,\cite{Loudon-book-1964,strach-prb-1998,Kranert-prl-2016} unaffected by volume,\cite{Li-2015-Nanoscale} dimensionality\cite{ribeiro-prb-2014,Wu-2018-CSR} and dielectric environment.\cite{Li-2015-Nanoscale,Lin-2019-NC} Moreover, for atomically-thin ALM flakes, the layer-number dependent e-phn/e-pht couplings\cite{Luo-2015-NC,Zhang-2018-SciAdv,Hwangbo-2021-NN,Qiao-2014-NC,Zhao-2021-NC,Song-2022-Nature,Zhang-2025-SciAdv} and optical anisotropy\cite{Qiao-2014-NC,Zhao-2021-NC,Song-2022-Nature} further complicate the interpretation of their ARPR intensity. Collectively, these factors impede a comprehensive understanding of ARPR intensity in ALMs, limiting systematic investigations of their optical anisotropy and e-phn coupling via ARPR spectroscopy. How to fully understand and quantitatively predict the ARPR intensity of ALMs flakes from atomically-thin layers to the bulk limit is a major challenge in this field.

In this work, we employ BP and Td-WTe$_2$ as model systems to investigate Raman scattering process in ALMs for quantitatively predicting their ARPR intensity profiles on various SiO$_2$/Si substrate, spanning from atomically-thin layers to bulk limit. The light propagation inside ALM flakes modulated from birefringence, linear dichroism and multilayer interference effects in air/ALMs/SiO$_2$/Si multilayers are rigorously modeled with the experimentally-determined complex refractive indexes along distinct crystallographic axes. We introduced the intrinsic Raman tensors (${\rm \textit{\textbf{R}}^{int}}$) for the Raman scattering event occurring inside ALMs, which allows for a complete understanding and quantitative prediction of the intricate dependence of ARPR intensity on $d_{\rm ALM}$, dielectric structures with different SiO$_2$ layer thickness ($d_{\rm SiO_2}$) and excitation wavelengths for thick BP and four-layer Td-WTe$_2$ (4L Td-WTe$_2$) flakes. Contour plots of the elements of effective Raman tensors (${\rm \textit{\textbf{R}}^{eff}}$) as functions of the thickness of BP flakes ($d_{\rm BP}$) and $d_{\rm SiO_2}$ for BP flakes and of $d_{\rm SiO_2}$ for 4L Td-WTe$_2$ are generated to quantitatively predict the corresponding ARPR intensity on arbitrary SiO$_2$/Si substrates without fitting parameters. This work provides a comprehensive insight into Raman scattering in ALM flakes and significantly broadens the scope for understanding their optical anisotropy.

\section{Results and Discussion}
\subsection{Sample characterization and thickness-dependent ARPR intensity of BP flakes}

\begin{figure*}[!ht]
  {\includegraphics[width=\linewidth]{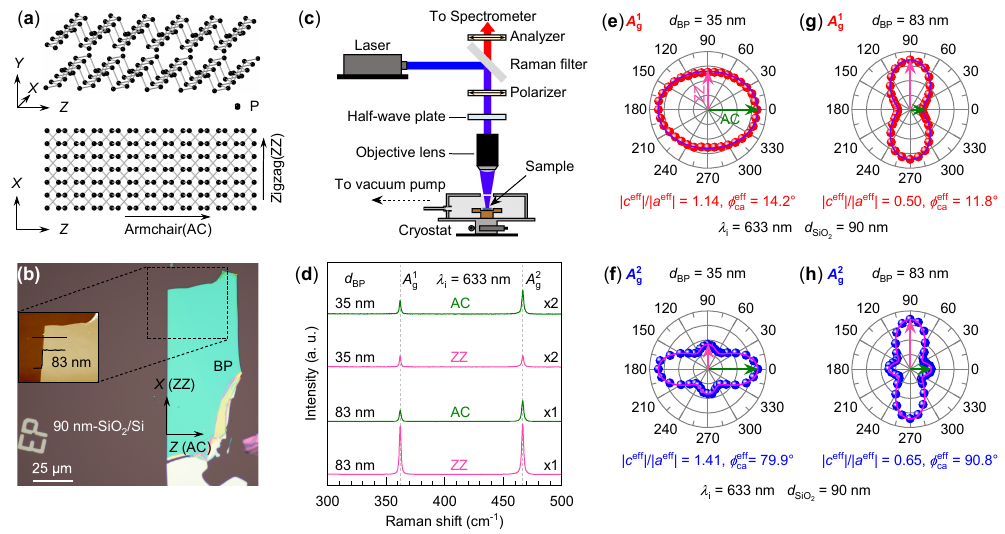}}
  \caption{a) Crystallographic structure of BP from the side and top views. b) Optical image of BP flake with $d_{\rm BP}$ = 83 nm measured by AFM (inset). c) Schematic diagram of ARPR spectroscopy setup with parallel polarization configuration. d) Raman spectra of BP flakes with $d_{\rm BP}$ = 35 nm and 83 nm on 90 nm-SiO$_2$/Si substrate, and the corresponding ARPR intensity profiles of the e,g) $A_{\rm g}^1$ and f,h) $A_{\rm g}^2$ modes of BP flakes excited by $\lambda_{\rm i} =$ 633 nm, where filled circles and solid line represent the experimental and the fitted ARPR intensity profiles, respectively. The fitted $|c^{\rm eff}|/|a^{\rm eff}|$ and $\mit\phi_{\rm ca}^{\rm eff}$ are indicated under each polar plot.}\label{Fig1}
\end{figure*}

We first use BP as a representative ALM to understand ARPR intensity in thick ALM flakes with bulk-like electronic structures. The orthorhombic symmetry (i.e., $D_{2h}$ symmetry) and puckered crystallographic structure of BP endow it with strong in-plane anisotropy (\textbf{Figure} \ref{Fig1}a). The $X$ and $Z$ axes were established in alignment with the in-plane zigzag (ZZ) and armchair (AC) directions,\cite{Lin-2020-SciBull} respectively, as illustrated in Figure \ref{Fig1}a. BP flakes were mechanically exfoliated onto SiO$_2$/Si substrates with varied $d_{\rm SiO_2}$. Figure \ref{Fig1}b shows the optical image of a BP flake with $d_{\rm BP} =$ 83 nm on 90 nm-SiO$_2$/Si substrate, as measured by atomic force microscopy (AFM). Earlier works conducted ARPR measurements by rotating ALM flakes \cite{Ribeiro-ACSNano-2015,Ling-Nanolett-2016,mao-jacs-2019,zou-nh-2021}. In this work, the Raman setup (Fig. \ref{Fig1}c) with a 20$\times$, low-NA (0.25) objective was used to measure the ARPR intensity of the BP flake under normal-incidence laser excitation on its basal plane using a 20$\times$, low-NA (0.25) objective. A polarizer placed upstream of the half-wave plate in the common optical path ensures a parallel polarization configuration, minimizing laser polarization alterations from system reflections/transmissions. An analyzer (aligned with the polarizer) before the spectrometer preserved the polarization state of selected Raman signals. Such configuration in Fig. \ref{Fig1}c guarantees the high quality of measured ARPR spectra. The angle ($\theta$) between $\textit{\textbf{e}}_{\rm i}$ ($\textit{\textbf{e}}_{\rm s}$) and the AC axis is controlled by a half-wave plate in the common optical path, with $\theta = 0^\circ$ for $\textit{\textbf{e}}_{\rm i}(\textit{\textbf{e}}_{\rm s})\parallel$ AC axis and $\theta = 90^\circ$ for $\textit{\textbf{e}}_{\rm i}(\textit{\textbf{e}}_{\rm s})\parallel$ ZZ axis. In this case, $\textit{\textbf{e}}_{\rm i}$ = $\textit{\textbf{e}}_{\rm s}$ = (sin$\theta$, 0, cos$\theta$)$^{\rm T}$. BP flakes were placed in a vacuum chamber during ARPR measurements to prevent oxidation.

Figure \ref{Fig1}d plots the Raman spectra of BP flakes with $d_{\rm BP}$ = 35 nm and 83 nm for $\textit{\textbf{e}}_{\rm i}(\textit{\textbf{e}}_{\rm s})\parallel$ AC and $\textit{\textbf{e}}_{\rm i}(\textit{\textbf{e}}_{\rm s})\parallel$ ZZ axes. The two typical Raman modes, i.e., $A_{\rm g}^1$ and $A_{\rm g}^2$ modes of BP flakes, are observed at 362 cm$^{-1}$ and 466 cm$^{-1}$, respectively. The Raman intensity ratio of $A_{\rm g}^1$ ($A_{\rm g}^2$) mode between $\textit{\textbf{e}}_{\rm i}(\textit{\textbf{e}}_{\rm s})\parallel$ ZZ axis and $\textit{\textbf{e}}_{\rm i}(\textit{\textbf{e}}_{\rm s})\parallel$ AC axis varies with $d_{\rm BP}$. The experimental ARPR intensity profiles for $A_{\rm g}^1$ and $A_{\rm g}^2$ modes of BP flakes were further demonstrated in Figure \ref{Fig1}e-h, which clearly show $d_{\rm BP}$ dependence. The Raman tensor for the $A_{\rm g}$ mode of BP flakes is as follows:

    \begin{equation}
    \textit{\textbf{R}}(A_{\rm g})=\left(
                       \begin{array}{ccc}
                         a & 0 & 0 \\
                         0 & b & 0 \\
                         0 & 0 & c \\
                       \end{array} \right)=\left(
                       \begin{array}{ccc}
                         |a|{\rm e}^{{\rm i}\phi_a} & 0 & 0 \\
                         0 & |b|{\rm e}^{{\rm i}\phi_b} & 0 \\
                         0 & 0 & |c|{\rm e}^{{\rm i}\phi_c} \\
                       \end{array}
                     \right).
    \label{tensor_BP}
    \end{equation}

\noindent For the normal incidence onto the basal plane, only the in-plane tensor elements $a$ and $c$ contribute to ARPR intensity. If considering effective Raman tensor ${\rm \textit{\textbf{R}}^{eff}}$ with complex tensor elements, i.e.,$a^{\rm eff} = |a^{\rm eff}|{\rm e}^{{\rm i}\phi_a^{\rm eff}}$ and $c^{\rm eff} = |c^{\rm eff}|{\rm e}^{{\rm i}\phi_c^{\rm eff}}$ ($\mit\phi_{\rm ca}^{\rm eff}$ = $\mit\phi_c^{\rm eff}-\mit\phi_a^{\rm eff}$),\cite{Ribeiro-ACSNano-2015,Kim-nanoscale-2015,Mao-small-2016,Ling-Nanolett-2016,mao-jacs-2019,Choi2020NH,pimenta-pccp-2021} , one can connect the experimentally measured ARPR intensity with $\textit{\textbf{e}}_{\rm i}$ and $\textit{\textbf{e}}_{\rm s}$ by

\begin{equation}
\begin{aligned}
I&\propto|\textit{\textbf{e}}^{\rm T}_{\rm s}\cdot {\rm \textit{\textbf{R}}^{eff}}\cdot \textit{\textbf{e}}_{\rm i}|^2\\
&=|a^{\rm eff}|^2{\rm cos}^4\theta+|c^{\rm eff}|^2{\rm sin}^4\theta+2|a^{\rm eff}||c^{\rm eff}|{\rm sin}^2\theta{\rm cos}^2\theta\cos\mit\phi_{\rm ca}^{\rm eff}.
\label{Ieff}
\end{aligned}
\end{equation}

By fitting the ARPR intensity with the Equation (\ref{Ieff}), $|c^{\rm eff}|/|a^{\rm eff}|$ and $\phi_{\rm ca}^{\rm eff}$ can be obtained. As shown in Figure \ref{Fig1}e-h, the fitted $|c^{\rm eff}|/|a^{\rm eff}|$ and $\phi_{\rm ca}^{\rm eff}$ are different for $A_{\rm g}^1$ and $A_{\rm g}^2$ modes and vary with $d_{\rm BP}$ for each Raman mode. This contradicts the basic physical picture that Raman tensor is an inherent parameter for a crystal, as the electronic band structure of BP flakes with few tens of nanometers only exhibit slight variations with change in $d_{BP}$\cite{Higashitarumizu-2023-NatureNanotech,Das-2014-NanoLetters}. Due to the in-plane anisotropy of BP flakes, the fitted $|c^{\rm eff}|/|a^{\rm eff}|$ and $\phi^{\rm eff}_{\rm ca}$ involve interplay of various anisotropy effects (birefringence,\cite{Mao-small-2016} linear dichroism\cite{Ribeiro-ACSNano-2015}, anisotropic e-pht and e-phn couplings\cite{Ling-Nanolett-2016}) and multilayer interference effect, distinct for different Raman modes and sensitive to $d_{\rm BP}$, making it a challenge to predict the ARPR intensity profiles of BP flakes.

\subsection{Anisotropic effects of Raman scattering process inside BP flakes}

\begin{figure*}[!ht]
  {\includegraphics[width=\linewidth]{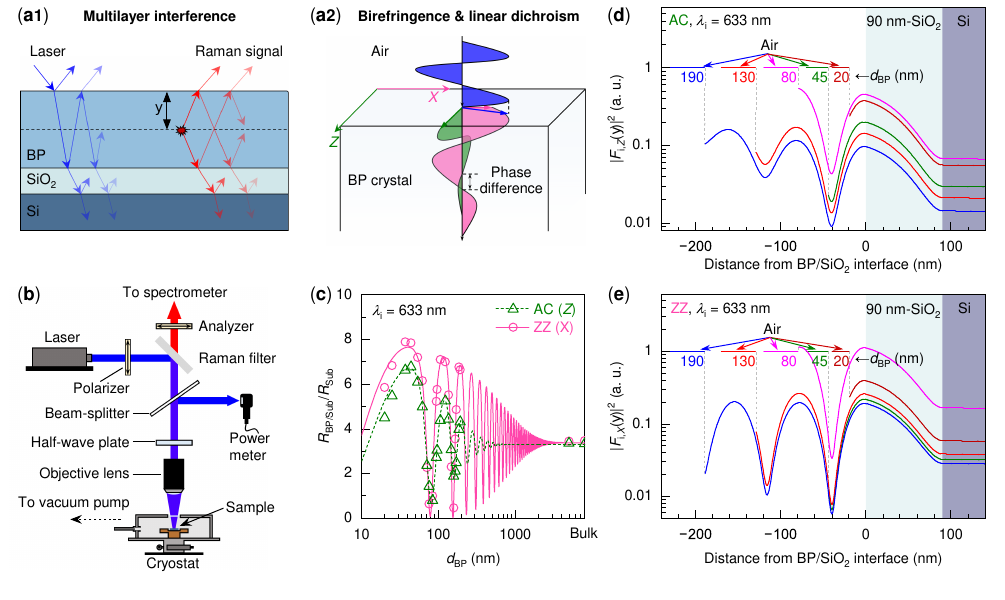}}
  \caption{a1) Schematic diagram of the multilayer interference of incident laser and scattered Raman signal inside multilayer structure, where the propagation paths of laser (blue lines) and scattered Raman signal (red lines) are presented separately. (Oblique incidence (scattering) for convenience). For clarity and simplicity, only a few representative reflection processes of the incident and scattered light are shown in the diagram. a2) Schematic illustration of the birefringence and linear dichroism in BP crystal. b) The setup for reflectance measurements. c) Experimental data (symbols) and fitted curves for the normalized reflectance of BP/90 nm-SiO$_2$/Si relative to the 90 nm-SiO$_2$/Si substrate as a function of $d_{\rm BP}$ along AC and ZZ axes at $\lambda_{\rm i}$ = 633 nm. The modulus square of the enhancement factors ($F_{{\rm i}} (y)$) for incident laser electric field inside BP/90 nm-SiO$_2$/Si multilayer structures at $\lambda_{\rm i}$ = 633 nm along d) AC and e) ZZ axes of BP flake, respectively.}\label{Fig2}
\end{figure*}

As described above, the complex tensor elements in ${\rm \textit{\textbf{R}}^{eff}}$ of the A$_g$ modes for BP flakes are dependent on $d_{BP}$, the dielectric substrate, and the excitation wavelength. Thus, to quantitatively understand the ambiguously dependence of ${\rm \textit{\textbf{R}}^{eff}}$ in BP flakes, all the anisotropy effects and multilayer interference effect in Raman scattering process should be considered, as depicted in \textbf{Figure} \ref{Fig2}a. BP flakes commonly deposited on SiO$_2$/Si substrate can generate a natural cavity due to the refractive index mismatch between BP flake and its underlying substrate\cite{MLLin-PRL-2025}, where partial reflections of incident (blue lines) and Raman scattered light (red lines) occur at air/BP, BP/SiO$_2$ and SiO$_2$/Si interfaces, as schematically depicted in Figure \ref{Fig2}a1 (see details in Figure S1 and Section 2, Supporting Information). As an ALM, BP flakes show evident birefringence and linear dichroism. As illustrated in Figure \ref{Fig2}a2, birefringence causes variation in phase velocities (i.e., phase difference) along the AC and ZZ axes of BP crystal, while linear dichroism leads to different penetration depths along these two axes.\cite{Lin-2020-SciBull} Thus, the polarizations and amplitudes of both the incident laser ($\textit{\textbf{e}}^\prime_{\rm i}(y)$) at the depth $y$ and Raman signal ($\textit{\textbf{e}}^\prime_{\rm s}(y)$) scattered into air from the depth $y$ inside BP flakes should be significantly different from  the experimentally-configured polarization vectors  $\textit{\textbf{e}}_{\rm i}$ and $\textit{\textbf{e}}_{\rm s}$ external to the ALM surface, respectively. Raman scattering occurring at $y$ in BP flakes is an inherent physical process driven by anisotropic e-pht and e-phn couplings, as described by the intrinsic Raman tensor ${\textit{\textbf{R}}^{\rm int}}$ (Section 1, Supporting Information). ${\textit{\textbf{R}}^{\rm int}}$ characterizes the polarizability changes caused by phonon atomic displacements during Raman scattering at the depth $y$. The Raman intensity $I(y)$ contributed from the depth $y$ inside BP flakes is given by

\begin{equation}
I(y)\propto|\textit{\textbf{e}}^{\prime {\rm T}}_{\rm s}(y)\cdot {\textit{\textbf{R}}^{\rm int}}\cdot \textit{\textbf{e}}^\prime_{\rm i}(y)|^2.
\label{intensity2}
\end{equation}

\noindent The modulation of incident and Raman scattered light by these anisotropic effects stems from varied complex refractive indexes along the $X$ ($\tilde{n}_{X}$) and $Z$ ($\tilde{n}_{Z}$) axes of BP flakes. Therefore, it is crucial to determine $\tilde{n}_{X}$ and $\tilde{n}_{Z}$. We use the setup in Figure \ref{Fig2}b to measure the reflectance along $X$ and $Z$ axes of flakes on SiO$_2$/Si substrate ($R_{\rm BP/Sub}$), in which the laser polarization direction is controlled by a half-wave plate in the common optical path. To eliminate the possible polarization-dependent light loss from objective and optical window of the vacuum chamber, $R_{\rm BP/Sub}$ was normalized by the reflectance of bare SiO$_2$/Si substrate (i.e., $R_{\rm Sub}$) without changing the focus status, i.e., $R_{\rm BP/Sub}/R_{\rm Sub}$. The measured $R_{\rm BP/Sub}/R_{\rm Sub}$ (triangles for $Z$ axis and circles for $X$ axis, respectively) for BP flakes on 90 nm-SiO$_2$/Si substrates with a variation of $d_{\rm BP}$ at $\lambda_{\rm i}$ = 633 nm are plotted in Figure \ref{Fig2}c. By fitting the experimental $R_{\rm BP/Sub}/R_{\rm Sub}$ with the normalized reflectance calculated via the transfer matrix method (TMM) (Section 2 and Section 3, Supporting Information), the fitted $\tilde{n}_Z$ and $\tilde{n}_X$ for BP flakes with $d_{\rm BP}$ ranging from 20 nm to $\sim$8 $\mu$m were obtained, as listed in \textbf{Table} \ref{Tab1}. The good agreement between the experimental and calculated results suggests that $\tilde{n}_Z$ and $\tilde{n}_X$ remain invariant for $d_{\rm BP}$ ranging from 20 nm to $\sim$8 $\mu$m, which should be ascribed to the analogous intrinsic band structures of thick BP flakes\cite{Higashitarumizu-2023-NatureNanotech,Das-2014-NanoLetters}, from 20 nm to bulk limit.

\begin{table}[t]
  \centering\caption{The experimentally-determined $\tilde{n}_{Z}$ and $\tilde{n}_{X}$ of BP flakes, and $|c^{\rm int}|/|a^{\rm int}|$, $\phi_{\rm ca}^{\rm int}$ for $A_{\rm g}$ modes in BP flakes on 90 nm-SiO$_2$/Si substrates that are averaged by the values fitted from BP flakes with $d_{\rm BP}$ = 54 nm and $\sim$5000 nm (bulk) at $\lambda_{\rm i}$ = 633 nm, 532 nm and 488 nm.}
  \resizebox{\linewidth}{!}{
    \begin{tabular}{ccccccc}
    \hline
    Wavelength[nm] & $\tilde{n}_{Z}$ & $\tilde{n}_{X}$ & $A_{\rm g}^{1}, |c^{\rm int}|/|a^{\rm int}|$ & $A_{\rm g}^{1}, \phi^{\rm int}_{\rm ca}$[$^\circ$] & $A_{\rm g}^{2}, |c^{\rm int}|/|a^{\rm int}|$ & $A_{\rm g}^{2}, \phi^{\rm int}_{\rm ca}$[$^\circ$]\\
    \hline
    633 &  4.01+0.38i $\pm$(0.08+0.03i) & 4.15+0.08i $\pm$(0.11+0.04i) & 1.16 & 19.5 & 1.48 & -68.8\\
    532 &  4.39+0.54i $\pm$(0.10+0.08i) & 4.48+0.13i $\pm$(0.11+0.06i) & 1.30 & 16.9 & 2.01 & -78.1\\
    488 &  4.50+1.02i $\pm$(0.11+0.08i) & 4.79+0.27i $\pm$(0.13+0.10i) & 1.02 & 63.2 & 2.61 & -63.4\\
    \hline
    \end{tabular}}
    \label{Tab1}
\end{table}

The modulations of birefringence, linear dichroism and multilayer interference on $\textit{\textbf{e}}^\prime_{\rm i}(y)$ and $\textit{\textbf{e}}^\prime_{\rm s}(y)$ coexist under above-bandgap excitation, e.g., 633 nm excitation. With the above-determined $\tilde{n}_Z$ and $\tilde{n}_X$, one can quantify these modulations by introducing the enhancement factor matrices of the incident laser ($\textit{\textbf{F}}_{\rm i}(y)$) and Raman signals ($\textit{\textbf{F}}_{\rm s}(y)$) at varied $y$, i.e., $\textit{\textbf{e}}^\prime_{\rm i}(y)$ = $\textit{\textbf{F}}_{\rm i}(y)\textit{\textbf{e}}_{\rm i}$ and $\textit{\textbf{e}}^\prime_{\rm s}(y)=\textit{\textbf{F}}_{\rm s}(y)\textit{\textbf{e}}_{\rm s}$, which take all the reflections or scattering processes of incident and scattered light in the multilayer structure into account. $\textit{\textbf{F}}_{\rm i(s)}(y)$ shows evident in-plane anisotropy and can be calculated using the TMM (see details in Section 2, Supporting Information),\cite{Li-2015-Nanoscale,Lin-2020-SciBull}

\begin{equation}
    \begin{aligned}
    &\textit{\textbf{F}}_{\rm i(s)}(y)={
    \left( \begin{array}{ccc}
    F_{{\rm i(s)},X}(y) & 0 & 0\\
    0 & 1 & 0\\
    0 & 0 & F_{{\rm i(s)},Z}(y)
    \end{array}
    \right)},
    \end{aligned}
    \label{Factor}
\end{equation}

\noindent where $F_{{\rm i(s)},X}(y)$ and $F_{{\rm i(s)},Z}(y)$ are respectively defined as the enhancement factors for incident laser (Raman scattered light) along $X$ and $Z$ axes. Figure \ref{Fig2}d,e illustrate the modulus square of $F_{{\rm i},Z}(y)$ and $F_{{\rm i},X}(y)$ for $\lambda_{\rm i}$ = 633 nm inside BP/90 nm-SiO$_2$/Si with typical $d_{\rm BP}$. Owing to multilayer interference effect in BP/SiO$_2$/Si structure, both $|F_{{\rm i},X}(y)|^2$ and $|F_{{\rm i},Z}(y)|^2$ exhibit cavity-like oscillatory distributions inside BP flakes, deviating from monotonic decrease. $|F_{{\rm i},X(Z)}(y)|^2$ also exhibits a pronounced $d_{\rm BP}$ dependence. In addition, the anisotropy in $\textit{\textbf{e}}^\prime_{\rm i}(y)$ is confirmed by different $|F_{{\rm i},Z}(y)|^2$ and $|F_{{\rm i},X}(y)|^2$. Since the Raman scattered light at each position $y$ of the BP flakes contributes to the whole Raman intensity, the measured Raman scattered intensity for a given phonon mode from BP flake is the integral of the Raman signal over $d_{\rm BP}$, as expressed below,

 \begin{equation}
    \begin{aligned}
  I\propto\int_{0}^{d_{\rm BP}}\left|\textit{\textbf{e}}^{\rm T}_{\rm s}\textit{\textbf{F}}^{\rm T}_{\rm s}(y)\cdot\textit{\textbf{R}}^{\rm int}\cdot \textit{\textbf{F}}_{\rm i}(y)\textit{\textbf{e}}_{\rm i}\right|^2 {\rm d}y.
    \label{Isum}
    \end{aligned}
 \end{equation}

\noindent Considering the nonzero tensor elements of ${\textit{\textbf{R}}^{\rm int}}$ for $A_{\rm g}$ modes of BP flakes, i.e., $R^{\rm int}_{xx}=a^{\rm int}=|a^{\rm int}|{\rm e}^{{\rm i}\phi_a^{\rm int}}$, $R^{\rm int}_{zz}=c^{\rm int}=|c^{\rm int}|{\rm e}^{{\rm i}\phi_c^{\rm int}}$ and $\phi_{\rm ca}^{\rm int}=\phi_c^{\rm int}-\phi_a^{\rm int}$, the Equation (\ref{Isum}) becomes:

\begin{equation}
    \begin{aligned}
    I\propto \int_{0}^{d_{\rm BP}}|&F_{{\rm i},X}(y)F_{{\rm s},X}(y)|a^{\rm int}| {\rm sin}^2\theta+F_{{\rm i},Z}(y)F_{{\rm s},Z}(y)|c^{\rm int}|{\rm e}^{{\rm i}\phi_{\rm ca}^{\rm int}}{\rm cos}^2\theta |^2 {\rm d}y.
    \label{IAg}
    \end{aligned}
\end{equation}

The above formalism clearly delineates the physical nature behind the ambiguous dependence of ARPR intensity profiles on $d_{\rm BP}$. The complicate ARPR intensity is influenced by the enhancement factors along both the ZZ and AC axes of BP flakes and the in-plane tensor elements ($a^{\rm int}$, $c^{\rm int}$) of $\textit{\textbf{R}}^{\rm int}$. In principle, once the $|c^{\rm int}|$/$|a^{\rm int}|$ ratio and $\phi_{\rm ca}^{\rm int}$ are determined, the ARPR intensity of BP flakes with any $d_{\rm BP}$ becomes predictable, regardless of the dielectric substrates.

\subsection{Determination of the intrinsic Raman tensor and prediction on ARPR intensity of BP flakes}

\begin{figure*}[!ht]
  {\includegraphics[width=\linewidth]{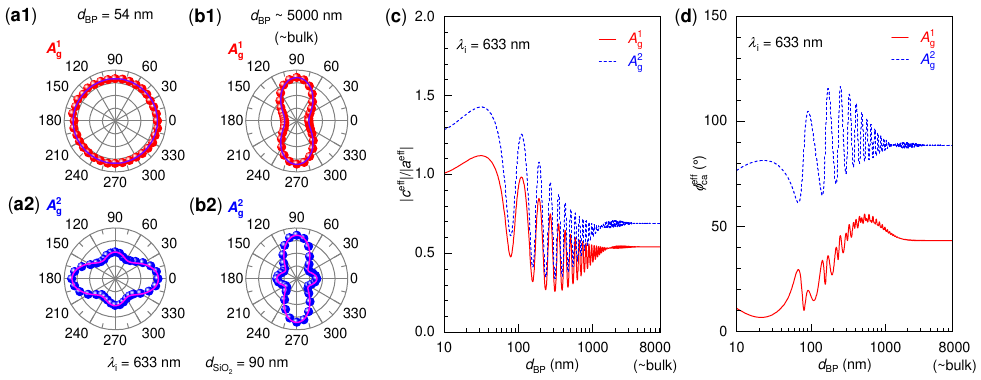}}
  \caption{Experimental (filled circles) and fitted (solid lines) ARPR intensity profiles at $\lambda_{\rm i}$ = 633 nm of $A_{\rm g}^1$ and $A_{\rm g}^2$ modes in BP/90 nm-SiO$_2$/Si with $d_{\rm BP}$ of a1,a2) 54 nm and b1,b2) $\sim$5000 nm (bulk), where the curves are fitted by Equation (\ref{IAg}). Predicted c) $|c^{\rm eff}|/|a^{\rm eff}|$ and d) $\mit\phi_{\rm ca}^{\rm eff}$ for $A_{\rm g}^1$ and $A_{\rm g}^2$ modes of BP flakes on 90 nm-SiO$_2$/Si substrate at $\lambda_{\rm i}$ = 633 nm.}\label{Fig3}
\end{figure*}

Using the determined $\tilde{n}_X$ and $\tilde{n}_Z$ from reflectance measurements, $F_{{\rm i(s)},X}(y)$ and $F_{{\rm i(s)},Z}(y)$ can be numerically computed. This allows the estimation of $|c^{\rm int}|/|a^{\rm int}|$ and $\phi^{\rm int}_{\rm ca}$ for the two modes in the BP flakes with specific thickness by fitting the ARPR intensity profiles with the Equation (\ref{IAg}). Using BP flakes with $d_{\rm BP}$=54 nm and 5000 nm as representative cases for thick BP flake (\textbf{Figure} \ref{Fig3}a) and bulk-like BP (\textbf{Figure} \ref{Fig3}b), we extract the $|c^{\rm int}|/|a^{\rm int}|$ and $\phi^{\rm int}_{\rm ca}$ for the $A^1_{\rm g}$ and $A^2_{\rm g}$ modes (Table S1, Supporting Information). The comparable magnitudes between flakes with $d_{\rm BP}$=54 nm and 5000 nm demonstrate that these parameters remain consistent for each Raman mode across different thick $d_{BP}$, reflecting their bulk-like electronic band structure\cite{Higashitarumizu-2023-NatureNanotech,Das-2014-NanoLetters}. This thickness invariance allows reliable estimation of these parameters using data from either $d_{BP}$, as estimated by averaging the values from 54 nm and 5000 nm samples, as shown in \textbf{Table} \ref{Tab1}. The $|c^{\rm int}|/|a^{\rm int}|$ for $A_{\rm g}^1$ and $A_{\rm g}^2$ modes are both larger than 1, indicating a more pronounced photon-phonon coupling mediated by electrons along the Z axis than along the X axis, as described by the interaction term $M_{\rm e-pht(s)}\cdot M_{\rm e-phn}\cdot M_{\rm e-pht(i)}$ (Section 1, Supporting Information), owing to the much larger light absorption along $Z$ axis.\cite{Qiao-2014-NC,Ling-Nanolett-2016} The larger $|c^{\rm int}|/|a^{\rm int}|$ ratio for the $A_{\rm g}^2$ mode ($\sim 1.48$) compared to that for the $A_{\rm g}^1$ mode ($\sim 1.16$) under 633 nm excitation suggests a greater disparity in $M_{\rm e-phn}$ between the $Z$ and $X$ axes for the $A_{\rm g}^2$ mode, attributed to the distinct vibrational behaviors of the two modes. Specifically, $A_{\rm g}^2$ mode predominantly exhibits in-plane vibrations along the $Z$ axis of BP and thus couples more significantly with the electronic transition dipole along in-plane $Z$ direction, whereas $A_{\rm g}^1$ mode mainly vibrating along out-of-plane direction couldn't efficiently couple with the electronic transition dipole along in-plane $Z$ direction. This is in line with the previous study.\cite{mao-jacs-2019} With the experimentally determined $\tilde{n}_X$, $\tilde{n}_Z$ and in-plane tensor elements of $\textit{\textbf{R}}^{\rm int}$, the ARPR intensities of thick BP flakes with arbitrary $d_{\rm BP}$ on any dielectric structure can be quantitatively predicted at $\lambda_{\rm i}$ = 633 nm using Equation \ref{IAg}.

Building on the insights into the modulations of Raman scattering process in BP flakes by birefringence, linear dichroism, multilayer interference and anisotropic e-pht (e-phn) coupling,\cite{Ling-Nanolett-2016,mao-jacs-2019} we aim to integrate all these effects to derive the formalism of the tensor elements of $\textit{\textbf{R}}^{\rm eff}$ to directly predict their ARPR intensity profile. By comparing Equation (\ref{IAg}) with Equation (\ref{Ieff}), the tensor elements in $\textit{\textbf{R}}^{\rm eff}$ can be derived as follows (see details in Section 4, Supporting Information),

\begin{equation}
    \begin{aligned}
    &\frac{|c^{\rm eff}|}{|a^{\rm eff}|}=\left(\frac{|c^{\rm int}|}{|a^{\rm int}|}\right)\cdot\left(\frac{\int^{d_{\rm BP}}_0 |A_Z|^2{\rm d}y}{\int^{d_{\rm BP}}_0 |A_X|^2{\rm d}y}\right),\\
    &{\rm cos}\phi^{\rm eff}_{\rm ca}=\frac{\left|\int^{d_{\rm BP}}_0 A_X^{\ast} A_Z{\rm d}y\right|}{\sqrt{\int^{d_{\rm BP}}_0 |A_X|^2{\rm d}y \int^{d_{\rm BP}}_0 |A_Z|^2{\rm d}y}}{\rm cos}(\mathit{\phi}^{\rm opt}_{\rm ca}+\mathit{\phi}^{\rm int}_{\rm ca}).
    \label{Rint}
    \end{aligned}
\end{equation}

\noindent Here, $A_{X(Z)}$ = $F_{{\rm i},X(Z)}(y)F_{{\rm s},X(Z)}(y)$, and $\mathit{\phi}^{\rm opt}_{\rm ca}$ = arg($\int^{d_{\rm BP}}_0 A_X^{\ast} A_Z{\rm d}y$) is defined as the principal argument of the phase difference arising from the optical anisotropy effects (i.e., birefringence, linear dichroism and multilayer interference) along $Z$ and $X$ axes of the BP flakes. Equation (\ref{Rint}) reveals the $d_{BP}$-dependent tensor elements in $\textit{\textbf{R}}^{\rm eff}$, clearly demonstrating the contributions from anisotropic e-pht and e-phn couplings as well as optical anisotropy effects. With the obtained $|c^{\rm int}|/|a^{\rm int}|$ and $\phi_{\rm ca}^{\rm int}$, we numerically calculated $|c^{\rm eff}|/|a^{\rm eff}|$ and $\phi^{\rm eff}_{\rm ca}$ for $A_{\rm g}^1$ and $A_{\rm g}^2$ modes in BP flakes on 90 nm-SiO$_2$/Si with $\lambda_{\rm i}$ = 633 nm, as illustrated in Figure \ref{Fig3}c,d. Both $|c^{\rm eff}|/|a^{\rm eff}|$ and $\phi^{\rm eff}_{\rm ca}$ are sensitive to $d_{\rm BP}$. With the derived tensor elements of $\textit{\textbf{R}}^{\rm eff}$, one can predict the ARPR intensity for BP flakes with Equation (\ref{Ieff}). Good agreements between the predicted and experimental ARPR intensity for $\lambda_{\rm i}$ = 633 nm are shown in Figure S2. This further confirms the consistent $|c^{\rm int}|/|a^{\rm int}|$ and $\phi^{\rm int}_{\rm ca}$ for thick BP flakes. The periodic variations of $|c^{\rm eff}|/|a^{\rm eff}|$ and $\phi^{\rm eff}_{\rm ca}$ of BP flakes give rise to periodic changes of the ARPR intensity shape. It is clear that the ambiguous dependencies of $|c^{\rm eff}|/|a^{\rm eff}|$ and $\phi^{\rm eff}_{\rm ca}$ on $d_{\rm BP}$ are the main reasons for the challenge in predicting ARPR intensity for BP flakes in previous studies.\cite{Ribeiro-ACSNano-2015,Kim-nanoscale-2015,Mao-small-2016,Ling-Nanolett-2016,Choi2020NH}

\begin{figure*}[!ht]
  {\includegraphics[width=\linewidth]{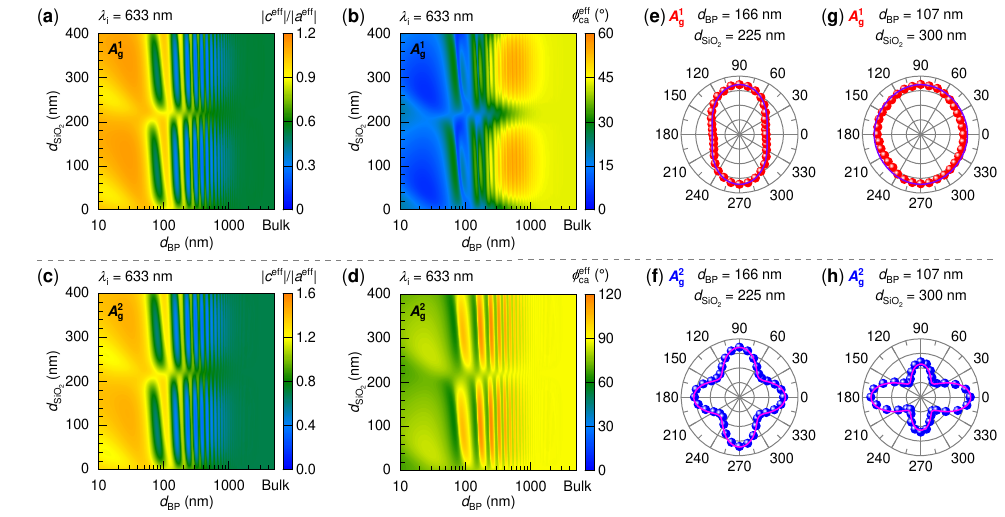}}
  \caption{Contour plots of $|c^{\rm eff}|/|a^{\rm eff}|$ and $\mit\phi_{\rm ca}^{\rm eff}$ for a), b) $A_{\rm g}^1$ and c), d) $A_{\rm g}^2$ modes with varied $d_{\rm BP}$ and $d_{\rm SiO_2}$ at $\lambda_{\rm i}$ = 633 nm. Experimental (filled circles) and predicted (solid lines) ARPR intensity profiles for $A_{\rm g}^1$ and $A_{\rm g}^2$ modes of BP flakes with e), f) $d_{\rm BP}$ = 166 nm on 225 nm-SiO$_2$/Si substrate and g), h) $d_{\rm BP}$ = 107 nm on 300 nm-SiO$_2$/Si substrate at $\lambda_{\rm i}$ = 633 nm.}\label{Fig4}
\end{figure*}

Owing to the evident optical interference for the BP/SiO$_2$/Si multilayer structure, $|c^{\rm eff}|/|a^{\rm eff}|$ and $\phi^{\rm eff}_{\rm ca}$ are also sensitive to $d_{\rm SiO_2}$. To extend the prediction of ARPR intensity profiles for BP flakes on various dielectric substrates, we plot the dependencies of $|c^{\rm eff}|/|a^{\rm eff}|$ and $\phi^{\rm eff}_{\rm ca}$ on $d_{\rm SiO_2}$ and $d_{\rm BP}$ for $A_{\rm g}^1$ and $A_{\rm g}^2$ modes at $\lambda_{\rm i}$ = 633 nm, as depicted in \textbf{Figure} \ref{Fig4}a-d. Both $|c^{\rm eff}|/|a^{\rm eff}|$ and $\phi^{\rm eff}_{\rm ca}$ show oscillatory variations with $d_{\rm SiO_2}$ and $d_{\rm BP}$ for thick BP flakes with $d_{\rm BP}$ ranging from 10 nm to 1000 nm, reflecting evident multilayer interference effect in this regime. The $|c^{\rm eff}|/|a^{\rm eff}|$ ratio of $A_{\rm g}^1$ mode consistently remains smaller than that of the $A_{\rm g}^2$ mode. Similar behavior are found for the $\phi^{\rm eff}_{\rm ca}$. With these predicted $|c^{\rm eff}|/|a^{\rm eff}|$ and $\phi^{\rm eff}_{\rm ca}$, the ARPR intensities for BP flakes on SiO$_2$/Si substrates with different $d_{\rm SiO_2}$ are predictable, as exemplified by the ARPR intensity of BP flake with $d_{\rm BP}$ = 166 nm on 225 nm-SiO$_2$/Si substrate (Figure \ref{Fig4}e,f) and BP flake with $d_{\rm BP}$ = 107 nm on 300 nm-SiO$_2$/Si (Figure \ref{Fig4}g,h). The predicted ARPR intensities well reproduce the measured ones, corroborating the validity of the strategy on predicting ARPR intensity  of BP flakes dependent on $d_{ALM}$ and dielectric substrates by the predicted effective complex Raman tensor elements, in which the dependence on dielectric substrates is hardly ever studied.

\begin{figure*}[!ht]
  {\includegraphics[width=\linewidth]{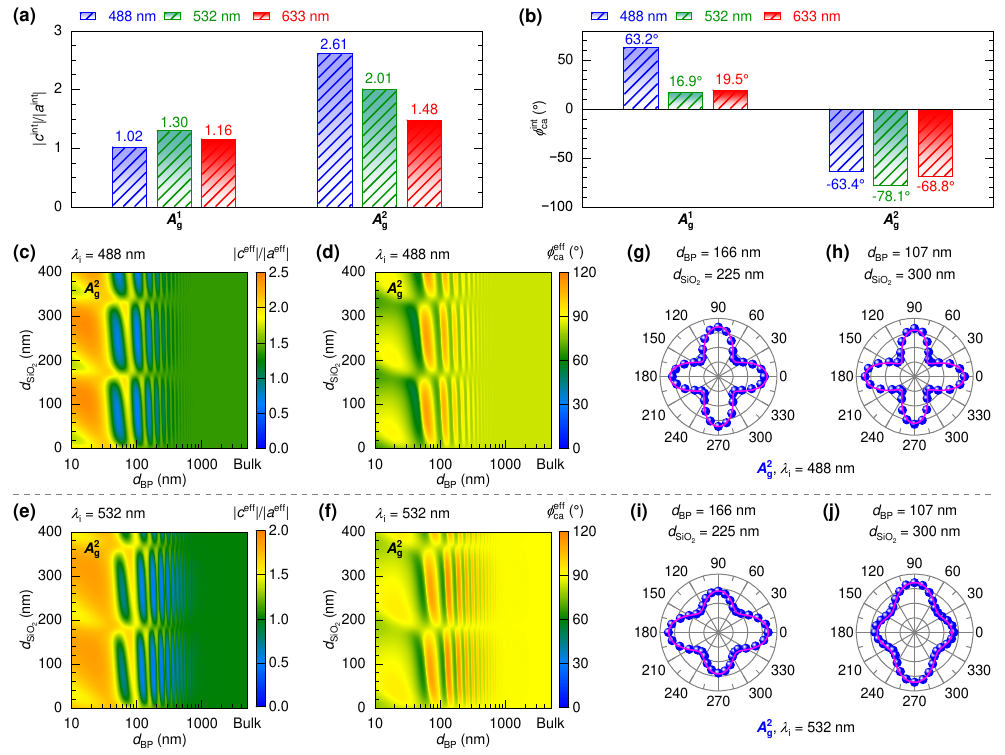}}
  \caption{Bar charts of the fitted averaged values of a) $c/a$ and b) $\phi^{\rm int}_{\rm ca}$ for $A_{\rm g}^1$ and $A_{\rm g}^2$ modes at $\lambda_{\rm i}$ of 488 nm (blue), 532 nm (green) and 633 nm (red), respectively. The values are indicated above the corresponding bars. Contour plots of $|c^{\rm eff}|/|a^{\rm eff}|$ and $\mit\phi_{\rm ca}^{\rm eff}$ for $A_{\rm g}^2$ mode with varied $d_{\rm BP}$ and $d_{\rm SiO_2}$ at $\lambda_{\rm i}$ of c), d) 488 nm and e), f) 532 nm. Experimental (filled circles) and predicted (solid lines) ARPR intensity profiles for $A_{\rm g}^2$ mode of BP flakes with $d_{\rm BP}$ = 166 nm on 225 nm-SiO$_2$/Si substrate and $d_{\rm BP}$ = 107 nm on 300 nm-SiO$_2$/Si substrate at $\lambda_{\rm i}$ of g), h) 488 nm and i), j) 532 nm.}\label{Fig5}
\end{figure*}

The above methodology to acquire complex refractive indexes ($\tilde{n}_X$, $\tilde{n}_Z$) and $\textit{\textbf{R}}^{\rm int}$ elements ($|c^{\rm int}|/|a^{\rm int}|$ and $\phi^{\rm int}_{\rm ca}$) of $A_{\rm g}^1$ and $A_{\rm g}^2$ modes for BP flakes were extended to other $\lambda_{\rm i}$, e.g., 488 nm and 532 nm, as illustrated in Figures S3-5 and in Table \ref{Tab1}. Notably, the 488 nm is at near-resonance excitation\cite{wang-jpcl-2018,mao-jacs-2019}. The averaged values of $|c^{\rm int}|/|a^{\rm int}|$ and $\phi_{\rm ca}^{\rm int}$ at three $\lambda_{\rm i}$ are statistically displayed in \textbf{Figure} \ref{Fig5}a,b. The results reveal that $|c^{\rm int}|/|a^{\rm int}|$ for $A_{\rm g}^2$ mode is larger than that for $A_{\rm g}^1$ mode at each specific $\lambda_{\rm i}$ of the three wavelengths, which is closely related to the varied vibrational characteristics of the phonon modes, as analyzed above for the 633 nm excitation. In addition, when $\lambda_{\rm i}$ = 488 nm, near-resonance excitation with the electronic band structure of BP induces a substantial $\phi_{\rm ca}^{\rm int}$ in $\textit{\textbf{R}}^{\rm int}$ for both $A_{\rm g}^1$ and $A_{\rm g}^2$ modes. With these parameters, one can calculate $|c^{\rm eff}|/|a^{\rm eff}|$ and $\phi^{\rm eff}_{\rm ca}$ in $\textit{\textbf{R}}^{\rm eff}$ for BP flakes on 90 nm-SiO$_2$/Si substrate with varied $d_{\rm BP}$ for $\lambda_{\rm i}$ = 488 nm and 532 nm (Figure S4e,f for 488 nm and Figure S5e,f for 532 nm, Supporting Information), which successfully reproduce the observed ARPR intensities of $A_{\rm g}^1$ and $A_{\rm g}^2$ modes for BP flakes without any additional fitting parameters (Figure S6 and Figure S7, Supporting Information). Furthermore, $|c^{\rm eff}|/|a^{\rm eff}|$ and $\phi^{\rm eff}_{\rm ca}$ also varies with $d_{\rm SiO_2}$ for $\lambda_{\rm i}$ = 488 nm and 532 nm, as presented in Figure \ref{Fig5}c-f (Figure S8a-d for $A_{\rm g}^{1}$ mode, Supporting Information). Based on these calculated parameters, the ARPR intensity of a BP flake on different dielectric substrates can be predicted, as exemplified by BP flakes with $d_{\rm BP}$ = 166 nm on 225nm-SiO$_2$/Si substrate and $d_{\rm BP}$ = 107 nm on 300 nm-SiO$_2$/Si at $\lambda_{\rm i}$ of 488 nm (Figure \ref{Fig5}g,h and Figure S8e,f in Supporting Information) and 532 nm (Figure \ref{Fig5}i,j and Figure S8g,h in Supporting Information). The predicted ARPR intensity profiles well reproduce the measured ones, suggesting the general validity of our proposed strategy to predict ARPR intensity profiles of ALM flakes with bulk-like electronic band structure under varied $\lambda_{\rm i}$ excitations by the experimentally-determined $\textit{\textbf{R}}^{\rm int}$, regardless of the intricate dependence of ARPR intensity profiles on $d_{\rm BP}$, dielectric substrates and excitation wavelengths. Furthermore, the good agreement between the predicted and experimental ARPR intensity under near-resonance excitation at 488 nm also implies that the proposed strategy is applicable to the resonance excitation case. Similar methodology can be extended to other ALM flakes for quantitative predictions of ARPR intensity profiles.

\subsection{Quantitatively predicting ARPR intensity of atomically-thin \texorpdfstring{Td-WTe$_2$}{}}

\begin{figure}[!ht]
  {\includegraphics[width=\linewidth]{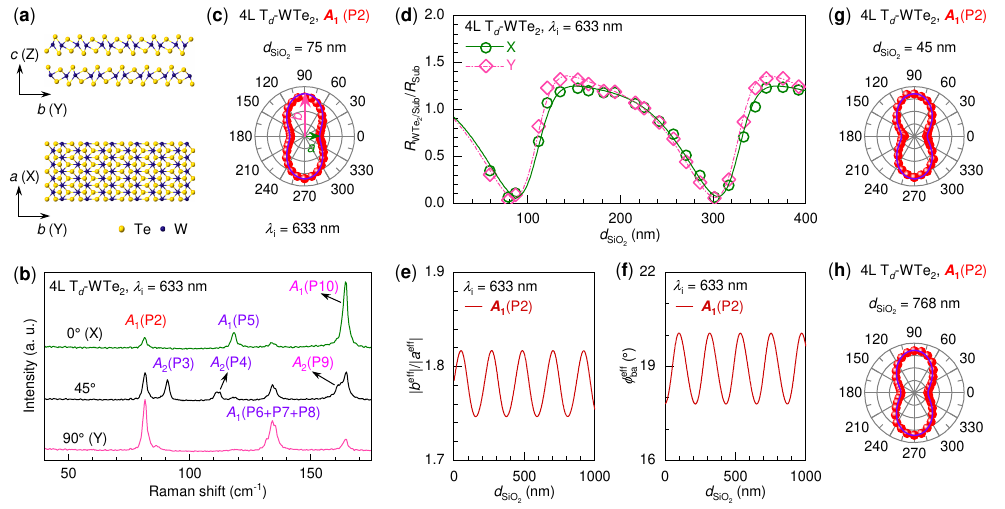}}
  \caption{a) Crystallographic structure of Td-WTe$_2$ from the side and top views. b) Polarized Raman spectra of 4L Td-WTe$_2$ on 75 nm-SiO$_2$/Si substrate at $\lambda_{\rm i}$ = 633 nm, when $\theta = 0^\circ$ ($e_{\rm i}~(e_{\rm s})\parallel X$), 45$^\circ$ and 90$^\circ$ ($e_{\rm i}~(e_{\rm s})\parallel Y$). The $A_1$ and $A_2$ modes are indicated. c) Experimental (filled circles) and fitted (solid line) ARPR intensity profiles of the P2 mode of 4L Td-WTe$_2$ on 75 nm-SiO$_2$/Si substrate at $\lambda_{\rm i}$ = 633 nm. d) Experimental data (Symbols) and fitted curves for the normalized reflectance of 4L Td-WTe$_2$/SiO$_2$/Si relative to the SiO$_2$/Si substrate as a function of $d_{\rm SiO_2}$ along $X$ and $Y$ axes at $\lambda_{\rm i}$ = 633 nm. The fitted $\tilde{n}$ for $X$ and $Y$ axes of Td-WTe$_2$ are $3.11+1.02{\rm i}$ and $3.31+0.95{\rm i}$, respectively (also see Table S1, Supporting Information). Predicted e) $|b^{\rm eff}|/|a^{\rm eff}|$ and f) $\mit\phi_{\rm ba}^{\rm eff}$ for the P2 mode in 4L Td-WTe$_2$ on SiO$_2$/Si substrate versus $d_{\rm SiO_2}$ at $\lambda_{\rm i}$ = 633 nm. Experimental (filled circles) and the correspondingly predicted (solid lines) ARPR intensity for the P2 mode in 4L Td-WTe$_2$ on g) 45 nm-SiO$_2$/Si and h) 768 nm-SiO$_2$/Si substrates at $\lambda_{\rm i}$ = 633 nm.}\label{Fig6}
\end{figure}

For atomically-thin ALM flakes, the layer-number dependent electronic band structures of few-layer ALMs results in layer-number dependent complex refractive index $\tilde{n}$ and Raman tensors. This invalidates the above strategies for determining $\tilde{n}$ along the in-plane axes by measuring the reflectance with a variation of $d_{ALM}$. To enable accurate prediction of ARPR in few-layer ALMs, we demonstrate an alternative approach using Td-WTe$_2$ flakes as a model system, circumventing the challenges associated with fabricating stable few-layer BP samples for most laboratories. Td-WTe$_2$ belongs to the orthorhombic crystal system with the space group Pmn2$_1$,\cite{Choi-2020-NM} which presents thermally stable phase and high sample quality down to few layers.\cite{Kim-2dm-2016,song-srep-2016,Choi-2020-NM} As shown in \textbf{Figure} \ref{Fig6}a, $a$ axis is defined along the W-W atomic chains, while another in-plane axis is always defined as $b$ axis.\cite{Kim-2dm-2016} We established the $X$ and $Y$ axes in alignment with the $a$ and $b$ directions of Td-WTe$_2$, according to the common definitions.\cite{Kim-2dm-2016,song-srep-2016} For the bulk Td-WTe$_2$, there are 33 Raman-active phonon modes at the $\Gamma$ point of the Brillion zone, but only phonon modes belong to $A_1$ and $A_2$ irreducible representations can be observed in the backscattering configuration.\cite{song-srep-2016} The corresponding Raman tensors are expressed as follows\cite{Kim-2dm-2016,song-srep-2016}

    \begin{equation}
    \textit{\textbf{R}}(A_1)=\left(
                       \begin{array}{ccc}
                         |a|{\rm e}^{{\rm i}\phi_a} & 0 & 0 \\
                         0 & |b|{\rm e}^{{\rm i}\phi_b} & 0 \\
                         0 & 0 & |c|{\rm e}^{{\rm i}\phi_c} \\
                       \end{array}
                     \right),                     \textit{\textbf{R}}(A_2)=\left(
                       \begin{array}{ccc}
                         0 & d & 0 \\
                         d & 0 & 0 \\
                         0 & 0 & 0 \\
                       \end{array}
                     \right).
    \label{tensor}
    \end{equation}

\noindent Thus, for $A_1$ modes, only $|a|{\rm e}^{{\rm i}\phi_a}$ and $|b|{\rm e}^{{\rm i}\phi_b}$ are involved at the normal incidence onto the basal plane. In addition, when $\textit{\textbf{e}}_{\rm i}\parallel\textit{\textbf{e}}_{\rm s}\parallel X$ ($Y$), i.e., $\theta=0^\circ (90^\circ)$, the corresponding Raman intensity $I$($X$)($I$($Y$)) is zero for $A_2$ modes.

Figure \ref{Fig6}b shows the Raman spectra of 4L Td-WTe$_2$ flake on SiO$_2$/Si substrate when $\theta = 0^\circ, 45^\circ$ and 90$^\circ$ at $\lambda_{\rm i}$ = 633 nm. Based on the analysis of the Raman tensor forms, the $A_1$ modes are observed in the Raman spectra when $\theta = 0^\circ$ and $\theta = 90^\circ$, while the $A_2$ modes are observed for $\theta = 45^\circ$. Additionally, the Raman tensor analysis also indicates that the ARPR intensity profiles of all the $A_2$ modes are similar to each other for Td-WTe$_2$ flake, independent of the tensor element values in $\textit{\textbf{R}}$. However, we found that $I$($Y$ axis)/$I$($X$ axis) of all the $A_1$ modes can be different from each other. For example, for 4L Td-WTe$_2$ flake on 75 nm-SiO$_2$/Si substrate, $I$($Y$)/$I$($X$)$>$1 for the P2 mode at around 80.7 cm$^{-1}$ (Figure \ref{Fig6}c), while $I$($Y$)/$I$($X$)$<$1 for the P10 mode at around 165 cm$^{-1}$.

To quantitatively predict ARPR intensity in 4L Td-WTe$_2$ flake, we demonstrate the approach using the P2 mode as an representative case. This requires prior determination of the anisotropic complex refractive indexes along $X$ ($\tilde{n}_{X}$) and $Y$ ($\tilde{n}_{Y}$) axes. Due to the distinct multilayer interference effects along $X$ and $Y$ axes, the reflectance along these two axes of 4L Td-WTe$_2$ flake is also sensitive to the dielectric substrate. This provides an additional approach to extract $\tilde{n}_{X}$ and $\tilde{n}_{Y}$ by measuring reflectance with varying $d_{\rm SiO_2}$ of SiO$_2$/Si substrates. \textbf{Figure} \ref{Fig6}d shows the $R_{\rm WTe_2/Sub}/R_{\rm Sub}$ for 4L Td-WTe$_2$ flake on SiO$_2$/Si substrates with $d_{\rm SiO_2}$ ranging from 60 nm to 390 nm for $\lambda_{\rm i}$ = 633 nm. By fitting the measured $R_{\rm WTe_2}/R_{\rm sub}$ with the calculated ones via TMM, the $\tilde{n}_{X}$ and $\tilde{n}_{Y}$ for 4L Td-WTe$_2$ flakes are obtained, as listed in Table S2 (Supporting Information). With these determined $\tilde{n}_{X}$ and $\tilde{n}_{Y}$, the corresponding enhancement factor matrices of the incident laser and Raman signals can be directly calculated. By fitting the ARPR intensities of the P2 mode of 4L Td-WTe$_2$ flake on 75 nm-SiO$_2$/Si substrate in Figure \ref{Fig6}c (replacing $c$ with $b$, and $Z$ axis with $Y$ axis in the formula for Td-WTe$_2$), the corresponding intrinsic in-plane tensor elements (amplitude ratio $|b^{\rm int}|/|a^{\rm int}|$ and phase difference $\phi^{\rm int}_{\rm ba}$  = $\phi_b^{\rm int}$-$\phi_a^{\rm int}$) in $\textit{\textbf{R}}^{\rm int}$ are determined, as listed in Table S2. Using the above parameters, we can predict the $|b^{\rm eff}|/|a^{\rm eff}|$ and $\phi^{\rm eff}_{\rm ba}$ for the P2 mode of 4L Td-WTe$_2$ flake on SiO$_2$/Si substrates with varied $d_{\rm SiO_2}$, as shown in Figure \ref{Fig6}e,f. Accordingly, the ARPR intensities of the P2 mode of 4L Td-WTe$_2$ flakes on other substrates can be also quantitatively predicted. For example, Figure \ref{Fig6}g,h illustrates the predicted ARPR intensities of the P2 mode of 4L Td-WTe$_2$ flakes on SiO$_2$/Si substrates with two $d_{\rm SiO_2}$ extremes, i.e., $d_{\rm SiO_2}$ = 45 nm and 768 nm, which are in line with the experimental results.

Due to the excitation-wavelength dependent complex refractive index and intrinsic Raman tensor (Section 1 of the Supporting Information), one should determine the complex refractive index and intrinsic Raman tensor for each excitation wavelength. The above proposed strategies are also extended to other wavelengths, e.g., $\lambda_{\rm i}$ = 532 nm (Figure S9, Supporting Information). Td-WTe$_2$ exhibits semimetal characteristics\cite{Kim-2dm-2016,song-srep-2016,Choi-2020-NM}, with no reported Raman resonance effects in the literature. Thus, we use a standard 532 nm laser to assess the universality of our quantitative prediction strategy for 4L Td-WTe$_2$ under different laser excitations. Using the derived $\textit{\textbf{R}}^{\rm int}$, the ARPR intensities for the P2 mode of 4L Td-WTe$_2$ under 532 nm excitation were successfully predicted on other substrates, as exemplified by the ones on SiO$_2$/Si substrates with $d_{\rm SiO_2}$ = 45 nm and 639 nm (Figure S9d,e). The predicted and experimental results are in good agreement with each other. Similar process can be applied to extract the $\textit{\textbf{R}}^{\rm int}$ for other Raman modes, while the complex refractive indexes extracted above by reflectance measurements can be applied to all Raman modes at a given excitation wavelength. The successful validations at two common excitation wavelengths (i.e., 633 nm and 532 nm) demonstrate the general validity of our proposed strategy to predict ARPR intensities of a specific few-layer ALM flake across different substrates and excitation wavelengths by the experimentally-determined $\textit{\textbf{R}}^{\rm int}$ elements.

Given the successful applications of our methodologies to both thick BP and 4L Td-WTe$_2$ flakes, despite their stark differences in symmetry, thickness, band topology, and optoelectronic properties, we contend that the detailed investigations of these two systems sufficiently demonstrate the universality of the proposed strategy across ALM flakes, spanning from atomically-thin layers to the bulk limit. Importantly, our strategy eliminates the need to consider whether the ALM flake satisfies the resonance condition for a specific excitation wavelength before performing ARPR spectroscopy of thick or few-layer ALM flakes. This unequivocally confirms that our universal method can quantitatively predict the ARPR intensity of any ALM flake, regardless of its thickness (from monolayer to bulk) or resonance condition.

\section{Conclusion}
In conclusion, this work achieves the quantitative prediction of ARPR intensity profiles in ALMs, ranging from atomically-thin layers to the bulk limit, which is determined by the interplay of birefringence, linear dichroism, multilayer interference of ALMs deposited on dielectric substrate and $\textit{\textbf{R}}^{\rm int}$. For ALM flakes with bulk-like electronic band structures and atomically-thin ALMs with layer-dependent electronic properties, we proposed strategies to quantitatively predict their ARPR intensities with the experimentally-determined complex refractive indexes and in-plane tensor elements ($|c^{\rm int}|/|a^{\rm int}|$, $|b^{\rm int}|/|a^{\rm int}|$, $\phi^{\rm int}_{\rm ca}$ and $\phi^{\rm int}_{\rm ba}$) of $\textit{\textbf{R}}^{\rm int}$. The tensor elements ($|c^{\rm eff}|/|a^{\rm eff}|$, $|b^{\rm eff}|/|a^{\rm eff}|$, $\phi^{\rm eff}_{\rm ca}$ and $\phi^{\rm eff}_{\rm ba}$) of $\textit{\textbf{R}}^{\rm eff}$ were also derived to predict ARPR intensity using a simplified equation. The methodology has been rigorously validated for atomically-thin and thick ALM flakes at different $\lambda_{\rm i}$ and substrates. The fact that our methodology successfully applies to both BP and Td-WTe$_2$ systems despite their stark differences in symmetry, band topology, and optoelectronic properties, which suggests its broad adaptability to ALMs with diverse characteristics. This study provides a comprehensive framework for overcoming challenges in predicting ARPR intensities of ALM flakes, thereby advancing the fundamental understanding of Raman scattering and optical anisotropy in ALM flakes.

\section{Experimental Section}
\threesubsection{Sample prepartion}
The BP crystals were synthesized by using a modified chemical vapor transport method with raw materials including amorphous red phosphorus, tin, and iodine.\cite{KOPF-jcg-2014} The as-prepared crystals were washed with acetone and ethanol to remove the surface adsorbed iodine. BP flakes were exfoliated from the bulk crystals onto polydimethylsiloxane (PDMS) sheets and were subsequently transferred onto SiO$_2$/Si substrates via the all-dry viscoelastic stamping method.\cite{Castellanos-2dm-2014} $d_{\rm BP}$ were measured by AFM under tapping mode. The ZZ and AC axes of BP flakes are determined by the relative Raman intensity ratio of the two $A_{\rm g}$ phonon modes, i.e., $I_{A^2_{\rm g}}({\rm AC})/I_{ A^1_{\rm g}}({\rm AC})>I_{A^2_{\rm g}}({\rm ZZ})/I_{\rm A^1_{\rm g}}({\rm ZZ})$ .\cite{Kim-nanoscale-2015,zou-nh-2021} And $d_{\rm SiO_2}$ of the SiO$_2$/Si substrate was determined by a spectroscopic ellipsometer.

4L Td-WTe$_2$ were prepared onto SiO$_2$/Si substrates via mechanical exfoliation and all-dry viscoelastic stamping method introduced above, in which the bulk Td-WTe$_2$ crystals were purchased from HQ graphene. We determine the layer number of Td-WTe$_2$ by the peak positions of the P2 mode around 80.7 cm$^{-1}$ and the wavenumber difference between the P2 mode and the P3 mode at $\sim$91 cm$^{-1}$, as they are sensitive to the layer number and can be utilized to determine the layer number of Td-WTe$_2$ flake according to the previous study.\cite{Kim-2dm-2016} In addition, the two in-plane axes ($a$($X$) and $b$($Y$) axes) of 4L Td-WTe$_2$ can be distinguished by the Raman intensity ratio of the P2 mode (or the P10 mode at $\sim$165 cm$^{-1}$) between $\textit{\textbf{e}}_{\rm i}(\textit{\textbf{e}}_{\rm s})\parallel X$ ($I(X)$) and $\textit{\textbf{e}}_{\rm i}(\textit{\textbf{e}}_{\rm s})\parallel Y$ ($I(Y)$) under the parallel polarization configuration, as $I(Y)/I(X)>1$ ($I(Y)/I(X)<1$) for the P2 mode (the P10 mode) under 633 nm excitation based on the previous work.\cite{Kim-2dm-2016}

\threesubsection{Raman measurements}
The Raman spectra were measured in backscattering geometry using a Jobin-Yvon HR-Evolution micro-Raman system equipped with an edge filter and a charge-coupled device detector. The excitation wavelengths are 633 nm from a He–Ne laser, 532 nm from a solid state laser, and 488 nm from a Ar$^+$ laser. For ARPR measurements, a 20$\times$ objective (NA = 0.25) was used to focus the incident light and collect the scattered Raman signal, ensuring near-normal incidence on the basal plane of the ALM flakes. A polarizer was placed in the common optical path of incident and Raman scattered light to achieve parallel polarization configuration, and an analyzer with polarization parallel to that of the polarizer was allocated before the spectrometer to ensure the preservation of the polarization state for the selected Raman signals. Zero-order half-wave plates at specified wavelengths of 633 nm, 532 nm and 488 nm were inserted in the common optical path between the objective lens and the polarizer to simultaneously vary the polarization directions of light when performing ARPR measurements under the three excitations, respectively. By rotating the fast axis of the zero-order half-wave plate with an angle of $\theta/2$, the polarizations of incident and Raman scattered light are rotated by $\theta$ relative to the $Z(X)$ axis of BP (Td-WTe$_2$) flakes synchronously. The laser power is less than 1.5 mW for 488 nm, 2 mW for 532 nm and 3 mW for 633 nm to avoid heating samples, respectively. To prevent possible degeneration during the measurements for investigating the intrinsic ARPR properties of the material, free from external environmental interference, the BP and Td-WTe$_2$ flakes were placed inside a vacuum chamber for ARPR measurements.

\threesubsection{Reflectance measurements}
The reflectance of ALM flakes along the interested crystallographic orientation can be measured by controlling the polarization direction of the incident laser. The lasers are the same as those used in Raman measurements. The polarization direction of the incident laser was purified by a linear polarizer. A beam splitter (Reflectance ($R$): Transmittance ($T$) = 45:55) was utilized to guide the laser onto the measured samples and separate the incident laser from the reflection light. Thus, we can measure the incident and reflection light intensity by a power meter. A zero-order half-wave plate was inserted into the common path of the incident laser and reflection light to vary the polarization directions of light when performing reflectance measurements. The laser polarization directions were controlled to be along the $X$ or $Z$ axes of BP flakes ($X$ or $Y$ axes of Td-WTe$_2$ flakes) to measure the corresponding complex refractive indexes, i.e., $\tilde{n}_{X}$ and $\tilde{n}_{Z}$ (or $\tilde{n}_{X}$ and $\tilde{n}_{Y}$). And an objective was utilized to focus the laser onto the measured samples. To avoid the possible degeneration of thin ALM flakes, the ALM flakes are implemented in a vacuum chamber. To calibrate the light loss induced by the used objective and optical window of the vacuum, the reflectance ($R_{\rm ALM/Sub}$) of the ALM flakes on SiO$_2$/Si substrates was normalized by the reflectance of the corresponding bare SiO$_2$/Si substrates near the BP flakes (i.e., $R_{\rm Sub}$) without changing the focus.
\vspace{1cm}

\raggedright
\medskip
\textbf{Supporting Information} \par 
Supporting Information is available from the Wiley Online Library or from the author.
\vspace{1cm}

\medskip
\textbf{Acknowledgements} \par 
\justifying
\noindent J.L. Xie and T. Liu contributed equally to this work. We acknowledge the support from the National Key Research and Development Program of China (Grant No. 2023YFA1407000), the Strategic Priority Research Program of CAS (Grant No. XDB0460000), National Natural Science Foundation of China (Grant Nos. 12322401, 12127807 and 12393832), CAS Key Research Program of Frontier Sciences (Grant No. ZDBS-LY-SLH004), Beijing Nova Program (Grant No. 20230484301), Youth Innovation Promotion Association, Chinese Academy of Sciences (No. 2023125) and CAS Project for Young Scientists in Basic Research (YSBR-026).
\vspace{1cm}
\medskip

%

\raggedright
\textbf{Conflict of Interest}\\
\noindent\justifying The authors declare no conflict of interest.
\vspace{1cm}

\raggedright
\textbf{Data Availability Statement}\\
\noindent\justifying The data that support the findings of this study are available on Research Gate data repository,\\ https://www.researchgate.net/publication/392559374.

\raggedright
\justifying
\bibliographystyle{MSP}

\end{document}